\documentstyle[aps, prl, twocolumn, epsf, floats]{revtex}

\begin{document}
\author{Y.-M. Vilk}
\address{Materials Science Division, Bldg. 223, 9700 S. Case Ave.}
\address{Argonne National Laboratory, Argonne IL 60439\\
$^{*}$ yvilk@hexi.msd.anl.gov}
\author{S. Allen, H. Touchette, S. Moukouri, L. Chen and A.-M.S. Tremblay}
\address{Centre de Recherche en Physique du Solide and D\'{e}partement de Physique}
\address{Universit\'{e} de Sherbrooke, Sherbrooke, Qu\'{e}bec, Canada J1K 2R1\\
tremblay@physique.usherb.ca}
\title{Attractive Hubbard model and single-particle pseudogap due to classical
pairing fluctuations in two dimensions}
\date{\today}
\maketitle

\begin{abstract}
{\it Abstract: }It is shown that in the two-dimensional attractive Hubbard
model, the mean-field phase transition is replaced by a renormalized
classical regime of fluctuations where a pseudogap opens up in the
single-particle spectral weight. It is argued that this pseudogap and
precursors of the ordered state quasiparticles can occur only in strongly
anisotropic quasi two-dimensional materials. This precursor phenomenon
differs from preformed local pairs. Furthermore, while critical
antiferromagnetic fluctuations would also lead to a pseudogap in the
repulsive model, there are some important differences between the
superconducting and magnetic pseudogap.
\end{abstract}

For several years, experimental evidence about the existence of gap-like
features in the normal phase of underdoped high-temperature superconductors
has been accumulating. In the last year or so, photoemission experiments
have convincingly shown that fermionic excitations do develop a pseudogap on
the Fermi surface\cite{ARPES}. This pseudogap, that forms well above the
superconducting transition temperature, suggests a normal-state precursor of
the superconducting gap. It has a similar angular dependence and basically
the same magnitude. Several explanations of this gap have been put forward.
They range from strong coupling effects, to magnetic effects, to preformed
local pair effects\cite{Preformed}. In this paper, we present a Quantum
Monte Carlo study of the attractive Hubbard model to show how precursors of
the superconducting quasiparticles may occur in the normal state, creating
pseudogap features in the single-particle weight $f\left( \omega \right)
A\left( {\bf k,}\omega \right) $ measured in Angular Resolved Photoemission
Experiments (ARPES). Pseudogap features caused by superconducting
fluctuations have a very long history, dating back to film studies in the
1970's.\cite{1970} Such fluctuations have been proposed again more recently
under the name of phase fluctuations\cite{PhaseFluct} as a cause of the
pseudogap. The pseudogap features discussed before were in the total density
of states. Conditions for a pseudogap in the momentum-dependent $f\left(
\omega \right) A\left( {\bf k,}\omega \right) $ are more stringent. These
conditions have been studied in the magnetic context\cite{Vilk} and
analogies with the superconducting case have also been pointed out.\cite
{LongPaper}

In this paper, we show that classical thermal fluctuations are sufficient to
obtain precursors of superconducting quasiparticles at arbitrarily small
coupling as long as a) the system is quasi two-dimensional b) the
superconducting correlation length $\xi $ grows faster with decreasing
temperature than the single-particle thermal de Broglie wavelength $\xi
_{th}=v_F/T.$ This last condition is always satisfied in the vicinity of
Kosterlitz-Thouless or mean-field transitions, although it is more stringent
than the condition $\xi >a$ ($a$ is the lattice constant) assumed by several
authors. The Physics of this effect is very different from that of the
so-called preformed local pairs that destroy the Fermi surface in any
dimension but only for sufficiently strong coupling.

The attractive Hubbard model, that leads to an $s-$wave state, is used here
because it is easier to simulate, allowing us to check the sufficient
conditions obtained analytically for the appearance of precursors of the
superconducting quasiparticles in the normal state. All simulations are done
for $U=-4t$ and units are chosen such that $t=1,a=1,$ $k_B=1,\hbar =1.$

At half-filling, where the chemical potential $\mu $ vanishes, the canonical
transformation $c_{i\downarrow }\rightarrow \left( -1\right)
^{i_{x}+i_{y}}c_{i\downarrow }^{\dagger }$ maps the attractive model onto
the repulsive one. The ${\bf q}=0,$ $s-$wave superconducting fluctuations
and the ${\bf q}=\left( \pi ,\pi \right) $ charge fluctuations are mapped
onto the three antiferromagnetic spin components of the repulsive model and
hence they are degenerate. Because of this degeneracy, the order parameter
at half-filling has $SO\left( 3\right) $ symmetry and hence, by the
Mermin-Wagner theorem, in two dimensions the phase transition cannot be at
finite temperature. At half-filling the transposition of the results for the
repulsive case shows that the pair structure factor $S_{\Delta }=\left(
\left\langle \Delta ^{\dagger }\Delta \right\rangle +\left\langle \Delta
\Delta ^{\dagger }\right\rangle \right) $ with $\Delta =\frac{1}{\sqrt{N}}%
\sum_{i=1}^{N}c_{i\uparrow }c_{i\downarrow }$ becomes size-dependent and
saturates at low temperature. This saturation signals a superconducting
ground state. Instead of a finite temperature phase transition, the sudden
rise of $S_{\Delta }$ as temperature decreases indicates a crossover to a
renormalized classical regime where the correlation length $\xi $ increases
exponentially. In an infinite system, a pseudogap opens up all the way from
the crossover temperature $T_{X}$ to zero temperature. The opening of this
pseudogap is signaled by a sharp decrease with temperature of the following
measure of the single-particle spectral weight near the Fermi energy: $%
\tilde{z}\left( T\right) =-2G\left( {\bf k}_{F},\beta /2\right) =\int \frac{%
d\omega }{2\pi }\frac{A\left( {\bf k}_{F},\omega \right) }{\cosh \left(
\beta \omega /2\right) }$.\cite{VT}

\begin{figure}%
%
\centerline{\epsfxsize 6cm \epsffile{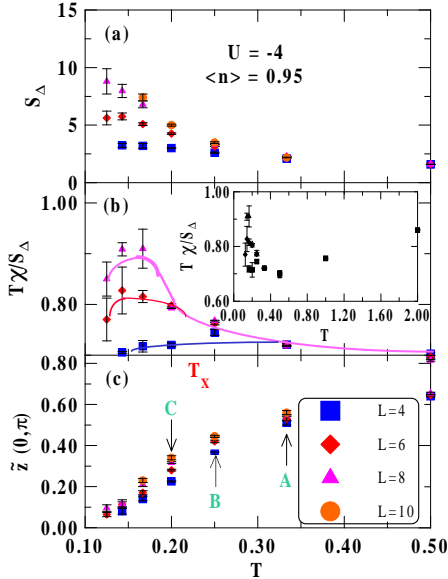}}%
%
\caption{(a) Pair structure factor. (b) Classical contribution. Lines are guides to the eye. (c) Weight at Fermi surface.}%
%
\label{fig1}%
%
\end{figure}%
%

Away from half-filling, the $SO\left( 3\right) $ symmetry is broken down to $%
SO\left( 2\right) $ by the finite chemical potential. The charge
fluctuations are suppressed compared with the superconducting ones. The
size-dependent data can be scaled like at half-filling with an exponentially
growing correlation length, showing that, for the accessible temperature
range, this is still an appropriate scaling. It is noteworthy that at the
crossover temperature $T_{X}$ the charge fluctuations are still comparable
to the superconducting ones. Despite the fact that a finite-temperature
Berezinski-Kosterlitz-Thouless (KT) transition is possible with an $SO\left(
2\right) $ symmetry, the rapid rise of superconducting correlations apparent
from Fig.1a is a crossover to a renormalized classical regime, not a
finite-temperature KT transition.\cite{MoreoScalapino} The fact that this is
a crossover is demonstrated by Fig.1b which shows the ratio of the classical
(zero Matsubara frequency) to quantum contribution to the pair structure
factor. We see that the classical contribution decreases as temperature
decreases (from $T=2$ to $T=1/2$ in the inset of Fig.1b) and then it starts
to rise again at lower temperature, signaling the onset of the renormalized
classical regime at $T_{X}\left( n\right) $. The downturn at lower
temperature indicates that as temperature becomes smaller than the minimal
energy of the pairing fluctuations in the given finite system, the finite
system returns back to a quantum regime and eventually $S_{\Delta }(L,q=0)$
saturates to its zero-temperature value in the superconducting {\it ground
state}. It is known from classical-spin simulations that the KT transition
is seen in finite-size systems only for much larger lattice sizes than those
accessible here, namely when at least several vortices can fit inside.

In the renormalized-classical regime that exists between $T_X$ and $T_{KT}$
one should observe features in the single-particle spectral weight that are
precursors of the superconducting ground state quasiparticles. This is
suggested by the drop in $\tilde{z}\left( T\right) $ in Fig.1c that
coincides with $T_X.$ To show the precursors more explicitly, consider in
Fig.2 the single-particle spectral weight $A\left( {\bf k}=\left( 0,\pi
\right) {\bf ;}\omega \right) $ obtained from analytical continuation of
Monte Carlo data using Maximum-entropy techniques. Temperature decreases
from $T=1/3$ to $1/5$ from the top to the bottom curve. These temperatures
correspond respectively to the points marked $A,B,C,$ in Fig.1c. Clearly,
the Fermi liquid quasiparticle that exists at high temperature, $T=1/3,$
starts to split into two superconducting-like quasiparticles around $T=1/4,$
opening a pseudogap at the Fermi level. The splitting at $T=1/4$ occurs in a
size-independent regime ($6\times 6$ near $8\times 8$) while at $T=1/5$ we
start to be in the size-dependent regime of Fig.1a. Note that ${\bf k}%
=\left( 0,\pi \right) $ is as close as possible to the Fermi surface. The
pseudogap should be larger for a point strictly on the Fermi surface.

\begin{figure}%
%
\centerline{\epsfxsize 6cm \epsffile{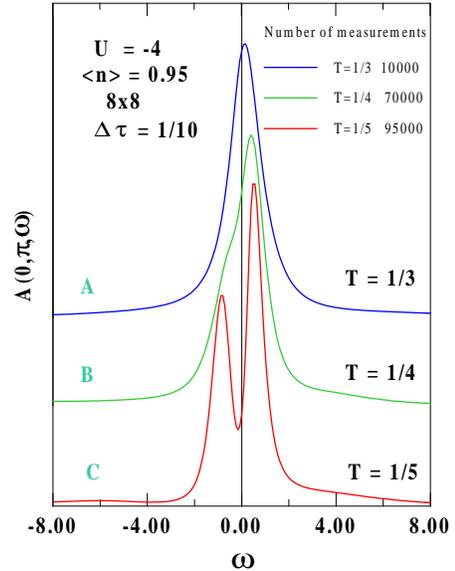}}%
%
\caption{Precursors near the zone edge}%
%
\label{fig2}%
%
\end{figure}%
%

The above results may be understood by analytical methods that have been
validated in detail in the half-filled case.\cite{LongPaper}\cite{VT}
Assuming some effective coupling constant $g$ between quasiparticles and
pairing fluctuations, one can obtain for the classical contribution to the
spectral weight, $\left( T>\left| \omega +\tilde{\epsilon}_{-{\bf k}}\right|
>v_{F}\xi ^{-1}\right) $, 
\begin{equation}
A({\bf k},\omega )\cong \frac{-2\Sigma _{cl}^{\prime \prime }({\bf k},\omega
)}{\left( \omega -\tilde{\epsilon}_{{\bf k}}-\frac{g}{2\pi }\frac{T\ln \xi }{%
\omega +\tilde{\epsilon}_{-{\bf k}}}\right) ^{2}+\Sigma _{cl}^{\prime \prime
}({\bf k},\omega )^{2}}
\end{equation}
As long as $\xi $ increases exponentially, it is clear that $\Delta
^{2}\propto gT\ln \xi $ will be finite so that the spectral weight will have
a maximum at frequency given by a BCS-like dispersion $\omega ^{2}=\tilde{%
\epsilon}_{{\bf k}}^{2}+\Delta ^{2}$. Clearly the above effects can exist
even in the weak coupling limit. This has to be contrasted with the case of
preformed local pairs that appear when the coupling strength becomes of the
order of the bandwidth. The latter is, certainly, not the case in High-T$%
_{c} $ materials where the gap is about $30meV$ while the bandwidth is of
the order $W\approx 1eV$. Also, preformed local pairs can occur for $U>W$ in
arbitrary dimensions. Here however, phase space considerations for the
integral defining the renormalized classical contribution show that {\it in
three dimensions} one does not obtain the singular form $T\ln \xi /\left(
\omega +\tilde{\epsilon}_{-{\bf k}}\right) $ for $\sum^{\prime }$ that is
necessary to obtain finite-frequency precursor quasiparticles. Nevertheless
the imaginary part of $\Sigma $ in three dimensions is weakly
(logarithmically) divergent close to $T_{c}$ which leads to a pseudogap in
the total density of states $N(\omega )=\frac{1}{N}\sum_{{\bf k}}A({\bf k}%
,\omega )$. Note that approximations that include self-consistency in the
Green's functions without the corresponding vertex corrections (FLEX like
approximations) fail to reproduce the precursor of superconducting bands in
the $k-$ resolved spectral function $A({\bf k},\omega )$ \cite{LongPaper}.

Physically, precursor effects exist because in the Kosterlitz-Thouless
picture the magnitude of the order parameter is locally non-zero starting
below a crossover temperature $T_X$ that is larger than the transition
temperature $T_{KT}$. It is only the phase that is globally decorrelated
above $T_{KT}$. Another way to understand the precursor effects is that the
superfluid density and the gap are {\em finite} as $T\rightarrow
T_{K.T.}^{-} $ and, hence, a two peak structure in $A({\bf k}_F,\omega )$
exists even as the phase transition point is approached from the
low-temperature side. This two peak structure should not immediately
disappear when one increases the temperature slightly above $T_{KT}$. In
quasi-two dimensional systems these high energy precursors should persist.
By contrast, in an isotropic three-dimensional system, the gap vanishes at
the transition point and there is no finite frequency precursors even though
there may be a low-frequency depletion of the total density of states.\cite
{LongPaper}

It is tempting to speculate that the physical origin of the decrease of $%
T_{c}$ in the underdoped materials is similar to what happens in the
attractive Hubbard model due to the fact that $T_{KT}$ must go to zero when
the number of components of the order parameter is larger than $N>2$ ( $%
SO\left( 3\right) $ in the attractive model). However, at present the size
of the fluctuating regime of the phase transition in high-$T_{c}$ materials
is controversial. Recent infrared experiments\cite{Timusk} suggesting a
pseudogap even in the overdoped regime are consistent with the mechanism
discussed here. Clearly however, present ARPES experiments\cite{ARPES} show
large backgrounds that are unexplained by the above approach. Finally, let
us contrast precursor effects due to magnetic and superconducting
fluctuations. Our criterion for the appearance of a pseudogap shows that
even for the s-wave interaction the superconducting pseudogap should open up
from the zone edge (smaller $v_{F}$) as temperature decreases until the
whole Fermi surface is gapped. This is, obviously, even more so for the $d-$
wave like interaction, when one has nodes along the diagonal direction. In
the case of an antiferromagnetic pseudogap, the region near the hot spots is
gapped, but a whole segment of the Fermi surface near the diagonal direction
remains gapless even when one approaches a phase transition.

We thank H.-G. Mattutis for discussion on Monte Carlo methods. We thank the
Centre d'application du calcul parall\`{e}le de l'Universit\'{e} de
Sherbrooke (CACPUS) for access to an IBM-SP2. This work was partially
supported by the Natural Sciences and Engineering Research Council of Canada
(NSERC), the Fonds pour la formation de chercheurs et l'aide \`{a} la
recherche from the Government of Qu\'{e}bec (FCAR), and (for Y.M.V.) by the
National Science Foundation (Grant No. NSF-DMR-91-20000) through the Science
and Technology Center for Superconductivity and (for A.-M.S.T.) by the
Canadian Institute for Advanced Research (CIAR).


\begin{references}
\bibitem{ARPES}  A.G. Loeser {\it et al,} Science {\bf 273}, 325 (1996); H.
Ding {\it et al.} Nature {\bf 382}, 51 (1996).

\bibitem{Preformed}  M. Randeria {\it et al.} Phys. Rev. Lett. {\bf 69},
2001, (1992); R. Micnas {\it et al.}, J. Phys. (Paris) Colloq. {\bf 49},
C8-2221 (1988).{\bf \ }For a review, M. Randeria in {\it Bose Einstein
Condensation }Ed. A. Griffin {\it et al.} (Cambridge University Press, 1995).

\bibitem{1970}  E. Abrahams {\it et al.} Phys. Rev. B {\bf 1}, 208 (1970);
B.R. Patton, Phys. Rev. Lett. {\bf 27}, 1273 (1971). For a review, M.
Ausloos, and A.A. Varlamov, {\it Proceedings of NATO Advanced Study
Institute }(Trieste, Aug. 5-9 1996).

\bibitem{PhaseFluct}  V.J. Emery and S.A. Kivelson, Nature {\bf 374}, 434
(1995).

\bibitem{Vilk}  Y. M. Vilk. Phys. Rev. B. {\bf 55}, 3870 (1997).

\bibitem{LongPaper}  Y.M. Vilk and A.-M.S. Tremblay, J. Phys. (Paris) (in
press, Nov. 1997) cond-mat/9702188.

\bibitem{VT}  Y.M. Vilk and A.-M.S. Tremblay, Europhys. Lett. {\bf 33}, 159
(1996).

\bibitem{MoreoScalapino}  A. Moreo and D.J. Scalapino, Phys. Rev. Lett. {\bf %
66}, 946 (1991).

\bibitem{Timusk}  T. Startseva, T. Timusk, A.V. Puchkov, D.N. Basov, H.A.
Mook, T. Kimura, K.Kishio, cond-mat/9706145.
\end{references}
\end{document}